\documentclass[nopreprintline, 12pt]{elsarticle}

\usepackage{amssymb}

\usepackage{amsmath}
\usepackage{amsthm}
\usepackage{multirow}
\usepackage{array}
\usepackage{hyperref}

\journal{Nuclear Physics B}

\begin{document}

\begin{frontmatter}

%% Title, authors and addresses

%% use the tnoteref command within \title for footnotes;
%% use the tnotetext command for theassociated footnote;
%% use the fnref command within \author or \address for footnotes;
%% use the fntext command for theassociated footnote;
%% use the corref command within \author for corresponding author footnotes;
%% use the cortext command for theassociated footnote;
%% use the ead command for the email address,
%% and the form \ead[url] for the home page:
%% \title{Title\tnoteref{label1}}
%% \tnotetext[label1]{}
%% \author{Name\corref{cor1}\fnref{label2}}
%% \ead{email address}
%% \ead[url]{home page}
%% \fntext[label2]{}
%% \cortext[cor1]{}
%% \affiliation{organization={},
%%             addressline={},
%%             city={},
%%             postcode={},
%%             state={},
%%             country={}}
%% \fntext[label3]{}

\title{Model Checking Access Control Policies: A Case Study using Google Cloud IAM}

%% use optional labels to link authors explicitly to addresses:
%% \author[label1,label2]{}
%% \affiliation[label1]{organization={},
%%             addressline={},
%%             city={},
%%             postcode={},
%%             state={},
%%             country={}}
%%
%% \affiliation[label2]{organization={},
%%             addressline={},
%%             city={},
%%             postcode={},
%%             state={},
%%             country={}}

\author[inst1]{Antonios Gouglidis}

\affiliation[inst1]{organization={School of Computing and Communications},%Department and Organization
            addressline={Lancaster University}, 
            city={Lancaster}, \\
            postcode={LA1 4WA}, 
            state={Lancashire},
            country={U.K.}}

\author[inst1]{Anna Kagia}
\author[inst2]{Vincent C. Hu}

\affiliation[inst2]{organization={Computer Security Division, NIST},%Department and Organization
            addressline={100 Bureau Drive}, 
            city={Gaithersburg}, \\
            postcode={2089}, 
            state={Maryland},
            country={U.S.A.}}

\begin{abstract}
%% Text of abstract
Authoring access control policies is challenging and prone to misconfigurations. Access control policies must be conflict-free. Hence, administrators should identify discrepancies between policy specifications and their intended function to avoid violating security principles. This paper aims to demonstrate how to formally verify access control policies. Model checking is used to verify access control properties against policies supported by an access control model. The authors consider Google's Cloud Identity and Access Management (IAM) as a case study and follow NIST's guidelines to verify access control policies automatically. Automated verification using model checking can serve as a valuable tool and assist administrators in assessing the correctness of access control policies. This enables checking violations against security principles and performing security assessments of policies for compliance purposes. The authors demonstrate how to define Google's IAM underlying role-based access control (RBAC) model, specify its supported policies, and formally verify a set of properties through three examples.
\end{abstract}

%%Graphical abstract
%\begin{graphicalabstract}
%\includegraphics{grabs}
%\end{graphicalabstract}

%%Research highlights
%\begin{highlights}
%\item Research highlight 1
%\item Research highlight 2
%\end{highlights}

\begin{keyword}
%% keywords here, in the form: keyword \sep keyword
Role-based access control \sep access control \sep authorization \sep policy verification \sep temporal logic \sep NuSMV 
%% PACS codes here, in the form: \PACS code \sep code
%\PACS 0000 \sep 1111
%% MSC codes here, in the form: \MSC code \sep code
%% or \MSC[2008] code \sep code (2000 is the default)
%\MSC 0000 \sep 1111
\end{keyword}

\end{frontmatter}

%% \linenumbers

%% main text
\section{Introduction}
The objective of an access control system is to control and limit the actions or operations in a system that an authorized user or process can perform on a set of resources \cite{sandhu1994access, ferraiolo2003role}. Access control is the process that checks all requests to a system and takes a decision to grant or deny access based on a set of rules. This makes it an essential component in all computing systems. In recent years, Cloud services have rapidly grown, rendering Cloud computing a popular computing paradigm. It changed the way organizations obtain IT resources and reduced costs significantly. As a result, Cloud computing has received considerable attention from academia as well as industry. Access control in the Cloud poses significant security challenges, e.g., secure inter-operation~\cite{hu2020general}, and supporting security assessment of policies~\cite{9159048}.

Access control policies dictate who has what access to which resource and thus it is important that these policies are error-free throughout their lifecycle. However, in practice, policies often do not satisfy the desired security requirements, and flaws in their specification can remain hidden and cause observable harm when exploited. Indeed, \cite{hu2017verification} states that misconfigurations in access control policies are one of the main reasons for security and privacy breaches due to potential inconsistencies. To eliminate unwanted access control discrepancies, verifying and rigorously testing access control policies before enforcing them in an operational environment is necessary. Nevertheless, the correct specification of access control policies is challenging since it is difficult to identify discrepancies between policy rule specifications and their intended functions for ensuring no violation of access control security principles \cite{gouglidis2014security}.

Although the integrated tools provided by Cloud providers can check policies for errors, Cloud administrators have little control over the specification of security requirements that can be  formally verified in access control policies. We anticipate that having an automated technique to verify the correctness of access control policies against a set of desired security requirements would serve as a valuable tool for Cloud administrators. This may assist in promptly identifying issues in the existing policies and provide information on how to exploit them. In this paper, we use an existing Identity and Access Management (IAM) system (i.e., Google's Cloud IAM) as a case study to elaborate on how policies can be modeled and subsequently verified against a set of user-defined properties.

The main contributions of this paper are:

\begin{itemize}
    \item Demonstrate how we formally define the RBAC model of IAM based on the limited publicly available information.  
    
    \item Specify a transition system for the RBAC model and demonstrate how to specify access control policies and properties in temporal logic.

    \item Verify user-defined properties in policy examples provided by Google through the above methods.
\end{itemize}

In the rest of this paper, we review some of the related work in Section~\ref{sec:2}, define Google's Cloud IAM RBAC model in Section~\ref{sec:3}, specify a transition system for the defined RBAC model and relevant properties in Section~\ref{sec:4}, verify example policies in Section~\ref{sec:5}, and present concluding remarks in Section~\ref{sec:6}. 

\section{Related Work}
\label{sec:2}
Zhang et al., \cite{zhang2019access} described the main Cloud access control models for OpenStack, AWS, and Microsoft Azure Cloud platforms. They provided a formal specification of these access control models and extended them to include the capability of handling information and resource sharing across tenants. Power et al.,  \cite{power2009modelling} presented two formal models of the access policy language used within the AWS Cloud computing infrastructure. They followed a hybrid approach by using both the Z specification language and the Alloy modeling language to test multiple policy properties and generate and test candidate policies. 
Evangelidis et al., \cite{evangelidis2018performance} proposed a probabilistic verification scheme based on performance modeling and formal verification of Cloud-based auto-scaling policies. To demonstrate the applicability of their method, they used a validation process on Amazon EC2 and Microsoft Azure, considering two different Cloud service models, i.e., IaaS and PaaS. Others focused on the challenges faced by the Cloud computing growth and conducted comparison studies between popular Cloud service providers, e.g., \cite{sleit2013cloud} compared Amazon EC2 and Microsoft Azure regarding how they deal with the challenges of availability, resource scaling, data deletion, data lock-in, and data security. Tajadod et al., \cite{tajadod2012microsoft} compared the same platforms looking at the security of architecture and the application levels.

A number of papers address verification of access control policies and several techniques have been reported in \cite{hu2017verification, fisler2005verification, hu2008property, li2015evaluating, aqib2015analysis, hu2006assessment, hughes2008automated, hoang2009specifying}. Their objectives are to look at methods that can check the correctness of policies. In this paper, we demonstrate the application of a generic technique, following NIST's guidelines \cite{hu2017verification}, which can verify access control properties against policies supported by an access control model.

In addition to the aforementioned approaches, a few access control verification tools were developed \cite{fisler2005verification, hwang2010acpt, jayaraman2011automatic, alloy2012, martin2008assessing} to facilitate policy-testing, with Access Control Policy Tool (ACPT) \cite{hwang2010acpt} and Security Policy Tool (SPT) \cite{infobeyond2017} as representative examples. The NIST Computer Security Division developed ACPT in collaboration with the North Carolina State University and University of Arkansas\cite{acpt2019}, and it is an implementation of the verification method in \cite{hu2017verification}. Through a graphical user interface (GUI), it provides templates for composing access control policies and properties and verifying them using a symbolic model verification (SMV) checker, NuSMV \cite{cavada2014nuxmv}. Moreover, it provides a complete test suite generated by NIST's combinatorial testing tool ACTS \cite{nist-ct} and generates XACML policy outputs of the verified model. SPT provides the same fundamental functions as ACPT with an extension of adding advanced features as a commercial product  \cite{infobeyond2017}.

\section{The \texorpdfstring{$RBAC_{GCP}$}{RBAC GCP} Model}
\label{sec:3}
Cloud IAM is part of the Google Cloud Platform (GCP), allowing Cloud administrators to control users' access to resources. Hence, when enforcing a policy, an organization can meet its regulatory and business objectives \cite{googleb2020}. \textit{"Cloud IAM manages access control by defining who (identity) has what access (role) for which resource"} \cite{googleOverview2020}. A high-level description of the RBAC model used in Google's Cloud IAM is available. Although its formal definition is not provided, Google documents its main entities, relations, and main operations. We formally define an access control model for Cloud IAM by following publicly available information and specify it based on the ANSI INCITS 359-2012 RBAC  \cite{incits2017incits}, which provides a solid foundation for defining role-based models. The following sections provide formal definitions of the main elements and functionalities of the model. Henceforth, we refer to the GCP RBAC model as $RBAC_{GCP}$.

\subsection{Model Description}

\begin{figure*}[!b]
  \centering
  \includegraphics[width=\linewidth]{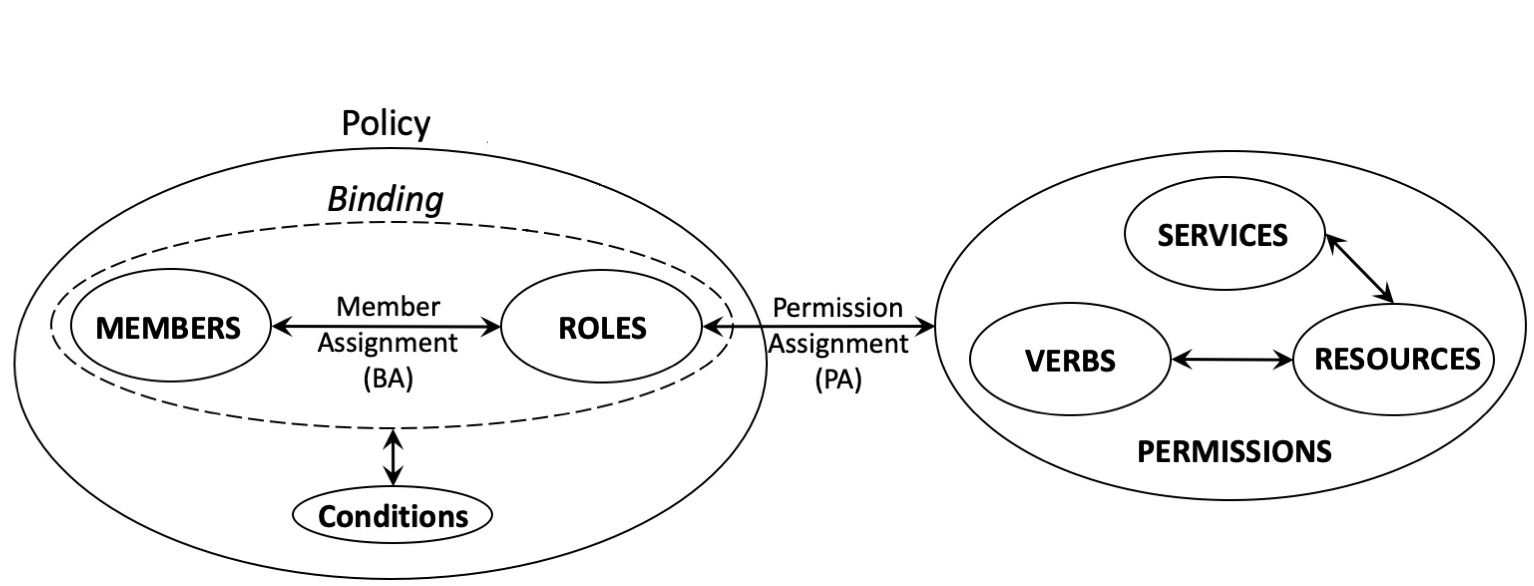}
  \caption{The $RBAC_{GCP}$ model.}
  \label{fig:1}
\end{figure*}

The $RBAC_{GCP}$ model consists of eight elements: MEMBERS, ROLES, PERMISSIONS, RESOURCES, SERVICES, VERBS, POLICIES, and CONDITIONS. It binds MEMBERS to ROLES and ROLES to PERMISSIONS instead of assigning PERMISSIONS directly to MEMBERS \cite{googleOverview2020}. Figure~\ref{fig:1} illustrates the relation of $RBAC_{GCP}$ elements. A MEMBER representing a human user or autonomous entity can access RESOURCES through a ROLE representing a job function described by a collection of PERMISSIONS. PERMISSIONS determine what VERBS (i.e., operations) are allowed on a system's RESOURCE (e.g., Compute Engine instances, Cloud Storage buckets). A POLICY is a collection of ROLE bindings, which bind one or more MEMBERS to individual ROLES. CONDITIONS assigned on ROLE bindings are logical expressions based on Google's Common Expression Language (CEL) and assigned on ROLE bindings.

Typically, in Cloud IAM, MEMBERS can be of the following type: Google account, Service account, Google group, G Suite domain, or Cloud Identity domain \cite{googleOverview2020}. ROLES can be Primitives, Predefined, or Custom. Primitives are the three concentric roles that have always existed in the GCP console: the Owner, Editor, and Viewer ROLES. The Owner ROLE contains the PERMISSIONS of the Editor, and the Editor ROLE includes the Viewer's PERMISSIONS. Google creates and maintains predefined roles and can provide granular access to specific GCP resources. Each product in the Google Cloud platform has its predefined role since different types of operations apply to different resources. A particular kind of role in Cloud IAM is Custom, which allow administrators to combine one or more PERMISSIONS and create unique ROLES that satisfy their organizations' needs when predefined ROLES are insufficient. Custom roles can only be granted within the Organization and cannot be used to grant PERMISSIONS on RESOURCES owned by a different Organization. Maintaining custom roles poses a challenge for administrators in creating potential security risks despite their flexibility. These ROLES are user-defined, therefore, not maintained by Google. Also, they are not automatically updated when an administrator adds new permissions, features, or services to the GCP \cite{google2009a}. Consequently, administrators must always keep up with the changes and ensure that any new functionality is consistent with the existing access control policies so as not to violate the security principles of the Organization. This task is challenging and can be highly complex and time-consuming \cite{googlef2020}.

\begin{figure}[!b]
  \centering
  \includegraphics[width=\linewidth]{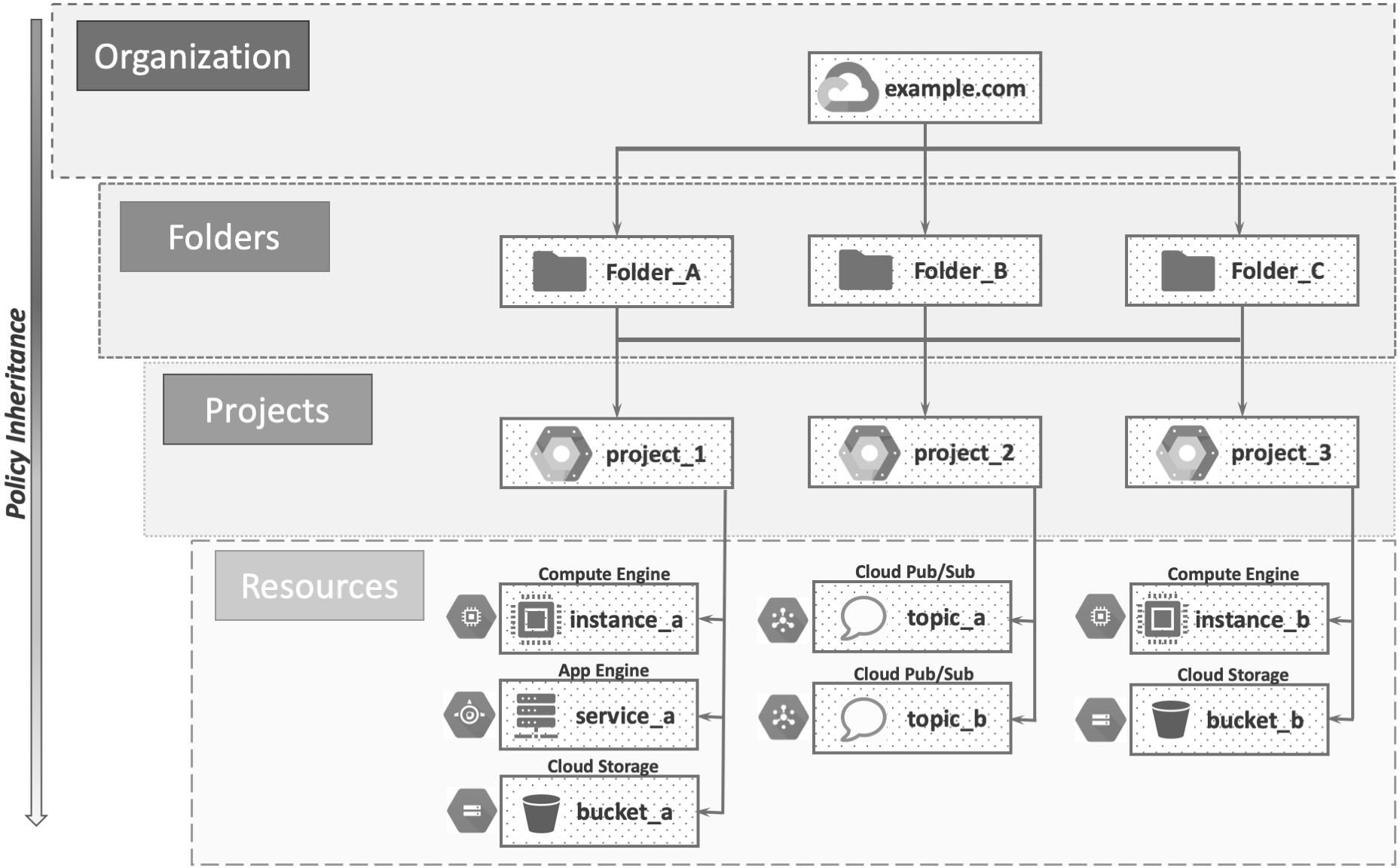}
  \caption{The Cloud IAM resource hierarchy (based on~\cite{google2020}).}
  \label{fig:2}
\end{figure}

PERMISSIONS in Cloud IAM are tuples  \textit{\textless service\textgreater, \textless resource\textgreater, \textless verb\textgreater} that describe using VERBS what OPERATIONS are allowed on a RESOURCE. A PERMISSION is defined per SERVICE and RESOURCE since every RESOURCE enables different OPERATIONS \cite{googleOverview2020}. For example, the PERMISSION \textit{"storage.buckets.create"} indicates creating a bucket in Cloud Storage is permitted for the storage service. RESOURCES are the fundamental components that comprise the GCP services, the Compute Engine instances (i.e., virtual machines), the App Engine services, the Cloud Storage buckets, and the Cloud Pub/Sub topics \cite{googleg2020}. RESOURCES in Cloud IAM are hierarchical, as shown in Figure~\ref{fig:2}. Projects are the children of the Folders, which are children of Organization, and the Resources are the descendants of Projects at the lowest level. Folders is an optional grouping mechanism.

\begin{figure}[!b]
  \centering
  \includegraphics[width=\linewidth]{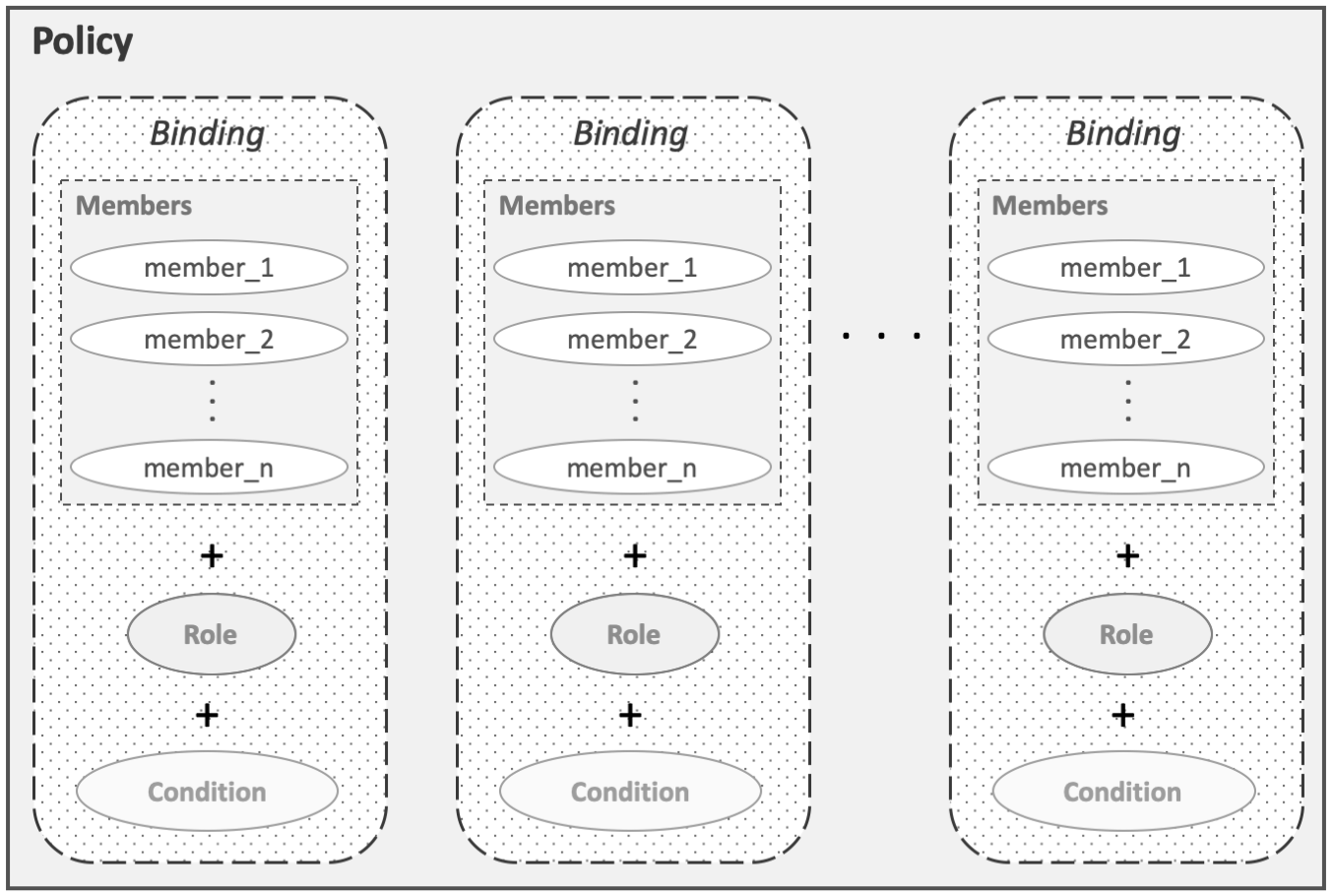}
  \caption{The Cloud IAM bindings (based on~\cite{googleOverview2020}).}
  \label{fig:3}
\end{figure}

POLICIES of Cloud IAM manage access to RESOURCES. A POLICY is a collection of statements that define the BINDING of ROLES and MEMBERS, as illustrated in Figure~\ref{fig:3} \cite{googleOverview2020}. BINDINGS can contain a CONDITION, an expression that includes one or multiple logic statements that evaluate various conditional attributes, which is optional, and each role BINDING may have only one. A BINDING without a CONDITION will always grant the ROLE to the specified MEMBERS. A BINDING is valid if a CONDITION is evaluated to TRUE. CONDITIONS provide constraints based either on the availability of a requested RESOURCE or on the situation of the access request. Examples for the former are the RESOURCE type and the RESOURCE name, and for the latter, the date/time of the request, the expected URL path, and the destination IP address. The enforcement of CONDITIONS can support attribute-based access control (ABAC) \cite{hu2013guide} to enhance the $RBAC_{GCP}$ model, allowing administrators to create more flexible and efficient access control policies. For instance, they can grant access to MEMBERS only during specified working hours and only for a specific RESOURCE type with the desired access level \cite{googlei2020}.

POLICIES are hierarchical and follow the same path as the RESOURCE hierarchy in Figure~\ref{fig:2}. That means that the administrator can set a policy at any level in the RESOURCE hierarchy (e.g.,  Organization, Folder, Project, Resource level), and the children's resources of that level can automatically inherit it. RESOURCES always inherit the POLICIES of the parent RESOURCE, and the inheritance is transitive through the hierarchical path. Therefore, RESOURCES inherit the POLICIES of the Project; Projects inherit the POLICIES of the Folder, and Folders inherit the Organization's POLICIES. At each level, the effective policies (i.e., in the presence of a hierarchy) are equal to the union of policies directly applied at the level and POLICIES inherited from its ancestors. For instance, a POLICY used in a Folder will also apply to Projects and RESOURCES under that Folder. Note that the POLICY hierarchy will change if the RESOURCE hierarchy is changed such that the PERMISSIONS that a child node inherited from its original parent will be lost and replaced by the PERMISSIONS set at the destination parent. $RBAC_{GCP}$ has no sessions. Instead, a ROLE remains dormant and not grantable if the respective SERVICE is not enabled. An administrator can use custom ROLES to enforce the principle of least privilege \cite{googlek2020}.

\subsection{Model Definition}
\label{sec:modeldefinition}

Following the notation used in the ANSI INCITS 359-2012 standard, we formally define the core $RBAC_{GCP}$ model as:

\begin{itemize}
    
    \item $\mathit{MEMBERS, ROLES, SERVICES, RESOURCES, VERBS,}$ $\mathit{CONDITIONS}$ are sets of members, roles, services, resources, verbs, and conditions, respectively;
    
    \item $\mathit{BINDING}$ is a binding, such as $\mathit{BINDING \subseteq MEMBERS \times ROLES \times CONDITIONS}$ is a many-to-many mapping relation of $\mathit{MEMBERS}$, $\mathit{ROLES}$ and $\mathit{CONDITIONS}$ assignment. $\mathit{CONDITIONS}$ are optional;
    
    \item $\mathit{PERMISSIONS = 2^{(SERVICES \times RESOURCES \times VERBS)}}$ is a set of permissions;
    
    \item $\mathit{PA \subseteq PERMISSIONS \times ROLES}$ is a many-to-many mapping of $\mathit{PERMISSIONS}$ to $\mathit{ROLES}$ assignment;
    
    \item $\mathit{POLICIES \subseteq 2^{BINDING}}$ is the set of policies, i.e., a single policy is a set of bindings.    
\end{itemize}

\section{Model and Properties Specification}
\label{sec:4}
This section elaborates on the model checking technique for verifying $RBAC_{GCP}$ policies. The process is compliant with NIST's guidelines \cite{hu2017verification}. Specifically, we define the $RBAC_{GCP}$ model using a transition system (TS). And verify example policies using temporal logic specifications for demonstration purposes.

\subsection{A Transition System for \texorpdfstring{$RBAC_{GCP}$}{RBAC GCP}}
\label{sec:TS}
Model checking is a formal verification technique that can be applied to verify the correctness of models and detect faults in model specifications. It takes a finite-state model and checks it against specified properties expressed using temporal modalities, linear temporal operators, and path quantifiers. To achieve this, we define access control rules in a transition system for the $RBAC_{GCP}$, as follows. 

\textbf{Definition 1.} An access control rule is an implication of $c \rightarrow d$, where constraint $c$ is a predicate expression of the form:

$\mathit{\left( \bigvee MEMBER = mbrs \right) \wedge \left( \bigvee ROLE = role \right) \wedge \left(\bigvee PERMISSION = prms \right)}$

$\mathit{\wedge \left( \bigvee RESOURCE=rscs \right)}$ which when $true$ implies the access control decision $d$, i.e.,  $\mathit{decision = Grant}$ or $\mathit{decision = Deny}$, where $\mathit{mbrs \in MEMBERS}$, $\mathit{role \in ROLES}$, $\mathit{prms \in PERMISSIONS}$, and $\mathit{rscs \in RESOURCES}$. The symbol of $\bigvee$ denotes that more than one formula may be present, e.g., $\bigvee MEMBER = mbrs$ could be $MEMBER = mbrs_1 \vee MEMBER = mbrs_2 \vee \ldots \vee MEMBER = mbrs_n$, where $mbrs_1, \ldots, mbrs_n \in MEMBERS$.

\textbf{Definition 2.} An $RBAC_{GCP}$ access control property $prop$ is an implication formula of $\forall \Box (c \rightarrow \forall \diamondsuit d)$, where $c$ is the cause and $d$ is the effect (response property pattern). Both $\Box$ and $\diamondsuit$ are elementary temporal modalities for "always" and "eventually", respectively, and $\forall$ means "for all paths" (Computation Tree Logic (CTL) semantics) \cite{CLARKE20011635}. 

\textbf{Definition 3.} The transition system $TS$ for the $RBAC_{GCP}$ model is expressed as a tuple $(S, Act, \delta, i_0)$ where:
\begin{itemize}
    \item $S$ is a set of system states, $S = \{Grant, Deny\}$;
    \item $Act$ is a set of actions, where
        $Act = \{ ( \left( \bigvee MEMBER = mbrs \right) \wedge \left( \bigvee ROLE = role \right) \wedge \left( \bigvee PERMISSION = prms \right) \wedge \left( \bigvee RESOURCE=rscs \right) \rightarrow decision = Grant), \ldots \} $
    \item $\delta$ is a transition relation, where $\delta : S \times Act \rightarrow S$;
    \item $i_0$ is the initial state, $i_0 = \{Deny\}$.
\end{itemize}

Access control rules define the system's behavior, which functions as the transition relation $\delta$ in $TS$. In other words, a transition system specifies how a system can evolve from one state to another when the transition relation is applied, i.e., an action $Act$ is performed on a state $S$ to bring the system to the next state of $S$. To verify $RBAC_{GCP}$ access control properties using a temporal logic formula, we say that model $TS$ satisfies $prop$ by $TS \models prop $ i.e., $TS \models \forall \Box (c \rightarrow \forall \diamondsuit d)$ from Definition 2.

\subsection{Specification of Properties}
\label{sec:CTLSpec}
The transition system describes the system's behavior, which can be used for verifying properties \cite{baier2008principles}. The verification shows if the access control policy is correctly specified and according to the security requirements. Specifically, model checking performs exhaustive testing of all behaviors of the model. It can verify if the defined properties hold or not throughout the model's behaviors (i.e., system states). In the $RBAC_{GCP}$ model, properties are expressed as (based on Definition 2; conditions are optional):

\begin{equation*}
\begin{aligned}
\forall \Box &( (MEMBER = m \wedge ROLE = r \wedge \\
             & PERMISSION = prms \wedge \\ 
             & RESOURCE = rsrc \wedge \\
             & CONDITION = value ) \rightarrow \\ 
             & \forall \diamondsuit (decision = Grant \vee Deny) ) 
\end{aligned}
\end{equation*}

Different specifications can be expressed depending on the values used in the predicates forming the property above. Consequently, we can define several different logical representations of the response pattern property using the same CTL formula.

\section{Verification of Example Policies}
\label{sec:5}
This section demonstrates using examples from Google’s Cloud IAM website  \cite{google2020} how to verify $RBAC_{GCP}$ policies. The examples show how the POLICY inheritance works in the Cloud IAM platform. We use these examples for their diversity in terms of used RESOURCES, MEMBER types, structural complexity, number of PERMISSIONS per ROLE, and level of a hierarchy of access control policy rules. The NuSMV code of all three examples are available on GitHub \cite{github2020}.

We assign values $m, r, prms, rsrc, value$ to the parameters $MEMBER$, $ROLE$, $PERMISSION$, $RESOURCE$, and $CONDITION$, respectively, following the CTL formula in Section 4.2 to specify properties. $CONDITION$ is optional and not used in the examples. The ANY value is introduced for all variables as a wild card. The response property is written as: $AG(c \rightarrow AF(d))$, where $G$ is an equivalent symbol used instead of $\Box$, and $F$ instead of $\diamondsuit$. $A$ represents the universal path quantifier $\forall$. So, we can rewrite access control properties in NuSMV as:
 
\begin{equation*}
\begin{aligned}
AG(( & MEMBER = m \& ROLE = r \& \\
     & PERMISSION = prms \& \\
    & RESOURCE = rsrc \&  \\
    & CONDITION = value) \rightarrow \\
    & AF(decision = Grant \mid Deny)).
\end{aligned}
\end{equation*}

The model checker creates all system model states and evaluates whether the policy model satisfies the specified properties. If it does, there are no errors from the output of NuSMV. Otherwise, a counterexample is generated, which details why the model fails to satisfy a property.

\subsection{Example 1: Cloud Pub/Sub}

\begin{figure}[!t]
  \centering
  \includegraphics[width=\linewidth]{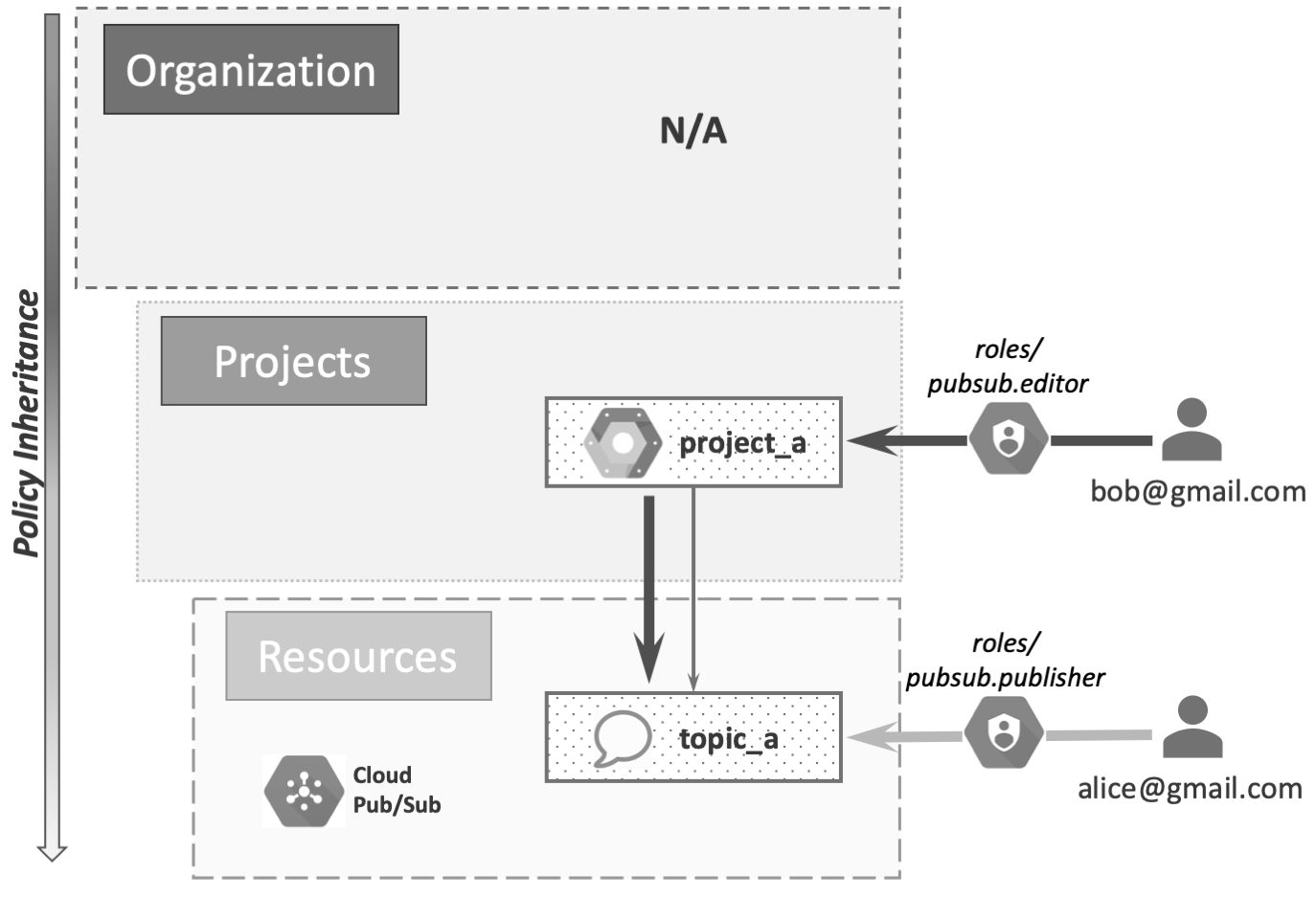}
  \caption{Example 1 - Cloud Pub/Sub (based on~\cite{google2020}).}
  \label{fig:5}
\end{figure}

The first example \cite{google2020} uses Cloud Pub/Sub RESOURCES, which are topics under a Project. As illustrated in Figure~\ref{fig:5}, topic\_a resides in project\_a. The Cloud IAM platform manages two Google accounts, i.e., $bob@gmail.com$ and $alice@gmail.com$. We assume that the POLICY $pl_1$ is set on $project\_a$ to assign the ROLE of Editor ($roles/pubsub.editor$) to $bob@gmail.com$ and POLICY $pl_2$ is set on $topic\_a$ to assign the ROLE of Publisher ($roles/pubsub.publisher$) to $alice@gmail.com$. Hence the two POLICIES that contain the rules are (based on Definition 1):

POLICY for \textit{bob@gmail.com}: 

\begin{equation*}
\begin{aligned}
pl_1: & MEMBER = "bob@gmail.com" \& \\
      & ROLE = "roles/pubsub.editor" \& \\
      & (PERMISSION = prms_1 \mid \dots \mid prms_n) \& \\
      & RESOURCE = "project\_a" \rightarrow \\
      & (decision = Grant)
\end{aligned}    
\end{equation*}

POLICY for \textit{alice@gmail.com}:

\begin{equation*}
\begin{aligned}
pl_2: & MEMBER = "alice@gmail.com" \& \\
      & ROLE = "roles/pubsub.publisher" \& \\
      & (PERMISSION = prms_1 \mid \dots \mid prms_n) \& \\
      & RESOURCE = "topic\_a" \rightarrow \\
      & (decision = Grant)
\end{aligned}
\end{equation*}

As RESOURCES always inherit the POLICIES of the parent RESOURCE, $topic\_a$ inherits the POLICY from $project\_a$. Hence, we introduce an additional POLICY $pl'_1$ for $topic\_a$ to assign the Editor ROLE $roles/pubsub.editor$ to $bob@gmail.com$, as follows:

\begin{equation*}
\begin{aligned}
pl_1': &MEMBER = "bob@gmail.com" \& ROLE = "roles/pubsub.editor" \& \\
       & (PERMISSION = prms_1 \mid \dots \mid prms_n) \& RESOURCE = "topic\_a" \rightarrow \\
       & (decision = Grant))
\end{aligned}
\end{equation*}

Ultimately the effective policy for $topic\_a$ will be the union of the POLICIES directly applied to $topic\_a$ and the POLICIES inherited from its ancestors. 

The respective NuSMV code for the POLICIES and the transition system is available in Example 1 on GitHub \cite{github2020}.
As a result, the ROLE assignments for each MEMBER per RESOURCE are shown in Table~\ref{ref:tbl1}.

\begin{table}[!t]
\caption{Example 1 - Authorized roles per member and resource.}
\label{ref:tbl1}
\centering
\begin{tabular}{|l|c|c|}
\hline
\multirow{2}{*}{\textbf{Resource}} & \multicolumn{2}{c|}{\textbf{Authorized role}}                                               \\ \cline{2-3} 
                                   & \multicolumn{1}{l|}{\textbf{bob@gmail.com}} & \multicolumn{1}{l|}{\textbf{alice@gmail.com}} \\ \hline \hline
project\_a                         & Editor                                      & No access                                     \\ \hline
topic\_a                           & Editor                                      & Publisher                                     \\ \hline
\end{tabular}
\end{table}

After expressing POLICIES and the $RBAC_{GCP}$ $TS$, the policy properties should be specified for verification in the model checker. When a specification is evaluated to be TRUE, there is no error to report, i.e., the specified property is satisfied by the model. On the other hand, when the specified property is not satisfied and evaluated to be FALSE, the model checker provides a counterexample to justify the result. For example, the NuSMV specification to check if $alice@gmail.com$ has the publisher ROLE for $project\_a$ at the $Projects$ hierarchy level is:

\begin{verbatim}
SPEC AG ((MEMBER = "alice@gmail.com") & 
(ROLE = "roles.pubsub.publisher") & 
(PERMISSION = ANY) & (RESOURCE = "project_a") -> AF decision = Grant) 
\end{verbatim}

The above will be evaluated to be FALSE since $alice@gmail.com$ is assigned to ROLE $Publisher$ on $topic\_a$, and according to $RBAC_{GCP}$ POLICY, she cannot access $project\_a$ because it resides in a higher level.
    
A NuSMV specification to check if $alice@gmail.com$ has the $pubsub.topics.publish$ PERMISSION on $project\_a$ at $Projects$  hierarchy level can be written:

\begin{verbatim}
SPEC AG ((MEMBER = "alice@gmail.com") &
(ROLE = ANY) & 
(PERMISSION = "pubsub.topics.publish") & 
(RESOURCE = "project_a") -> AF decision = Grant) 
\end{verbatim}

The above will be evaluated to be FALSE since $alice@gmail.com$ has the PERMISSION $pubsub.topics.publish$, for her $Publisher$ ROLE only on $topic\_a$, but not on $project\_a$ that resides on a higher level.
    
Lastly, to check if $alice@gmail.com$ has the PERMISSION $pubsub.topics.delete$ on $topic\_a$ at $Resources$ hierarchy level, we write:

\begin{verbatim}
SPEC AG ((MEMBER = "alice@gmail.com") & 
(ROLE = ANY) & 
(PERMISSION = "pubsub.topics.delete") & 
(RESOURCE = "topic_a") -> AF decision = Grant) 
\end{verbatim}
    
Although $alice@gmail.com$ has the ROLE $Publisher$ on $topic\_a$ she does not have the PERMISSION $pubsub.topics.delete$ since an assignment is missing between that ROLE and the PERMISSION; hence, it is evaluated to be FALSE.

In all three specifications, the result of the verification is $RBAC.decision = Deny$ without a next state, which indicates that they can never be satisfied, according to the $RBAC_{GCP}$ POLICIES. The model checker could not find any system state where the property verified to be TRUE for the access permission $Grant$ to happen. 

\subsection{Example 2: Cloud Storage}

\begin{figure}[!t]
  \centering
  \includegraphics[width=\linewidth]{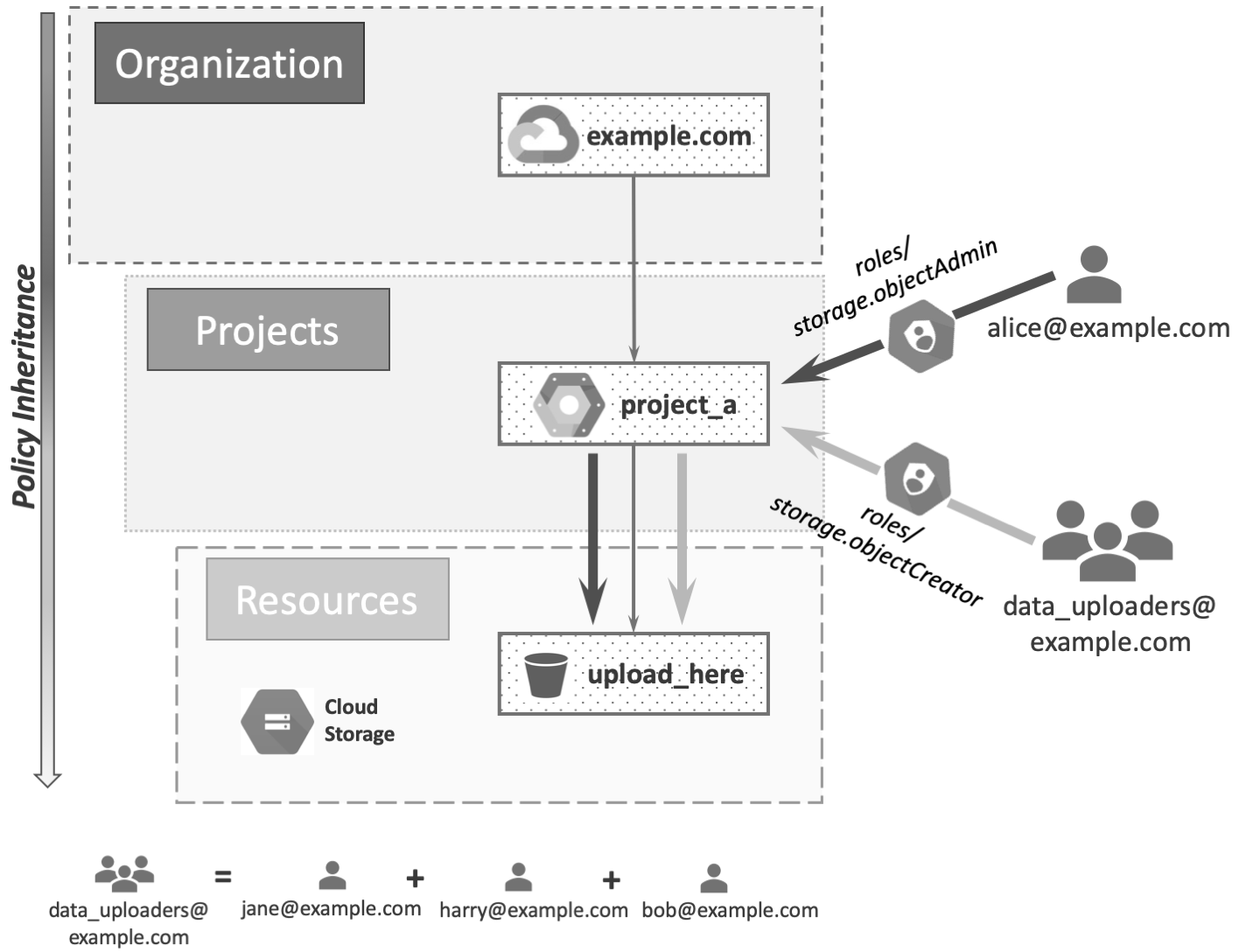}
  \caption{Example 2 - Cloud Storage (based on~\cite{google2020}).}
  \label{fig:6}
\end{figure}

The second example \cite{google2020} uses Cloud Storage RESOURCES called buckets. The bucket $upload\_here$ belongs to the Project $project\_a$ of the Organization $example.com$ and is used to store files uploaded from GCP users (see Figure~\ref{fig:6}). Many users can use the same bucket to upload files; thus, it requires that no user can delete any of the files uploaded by other users. However, the data processing expert should be able to gain or delete anyone's files. 

We assume that $alice@example.com$ is the Google account of the data processing expert and $data\_uploaders@example.com$ is the group account of users who upload files to the bucket. The group has three MEMBERS: $jane@example.com$, $harry@example.com$, and $bob@example.com$. To achieve the security requirements, a POLICY is set on $project\_a$ to assign the Storage Object Admin ROLE ($roles/storage.objectAdmin$) to $alice@example.com$, and a second POLICY is set on $project\_a$ to assign the Storage Object Creator ROLE ($roles/storage.objectCreator$) to $data\_uploaders@example.com$. These ROLES should allow $alice@example.com$ to upload or delete any object in any bucket in $project\_a$, while the MEMBERS of $data\_uploaders@example.com$ should be allowed to upload files. The two POLICIES will look as follows: 

POLICY for \textit{alice@example.com}: 

\begin{equation*}
\begin{aligned}
pl_1: & MEMBER = "alice@example.com" \& \\
      & ROLE = "roles/storage.objectAdmin" \& \\
      & (PERMISSION = prms_1 \mid \dots \mid prms_n) \& RESOURCE = "project\_a" \rightarrow \\
      & (decision = Grant)
\end{aligned}
\end{equation*}

POLICY for data \textit{data\_uploaders@example.com}:

\begin{equation*}
\begin{aligned}
pl_2: & (MEMBER = "data\_uploaders@example.com" \mid "jane@example.com" \mid \\
      & "bob@example.com" \mid "harry@example.com") \& \\
      & ROLE = "roles/storage.objectCreator" \& \\
      & (PERMISSION = prms_1 \mid \dots \mid prms_n) \& RESOURCE = "project\_a" \rightarrow \\
      & (decision = Grant)
\end{aligned}    
\end{equation*}

POLICY $pl_2$ applies to $project\_a$ for every group MEMBER, which assigns the Storage Object Creator ROLE to $jane@example.com$, $harry@example.com$ and $bob@example.com$, as well.

Bucket $upload\_here$ inherits POLICIES from its parent RESOURCE $project\_a$. POLICIES $pl_1$ and $pl_2$ will then be defined and populated to the transition system of the $RBAC_{GCP}$ model. Although the bucket has no defined POLICIES, these two POLICIES will apply on $upload\_here$ (due to hierarchy) such that the Storage Object Admin ROLE is assigned to $alice@example.com$ on $upload\_here$, and the Storage Object Creator ROLE is assigned to $data\_uploaders@example.com$ for $upload\_here$, as follows:

\begin{equation*}
\begin{aligned}
pl_1': & MEMBER = "alice@example.com" \& \\
       & ROLE = "roles/storage.objectAdmin" \& \\
       & (PERMISSION = prms_1 \mid \dots \mid prms_n) \& \\
       & RESOURCE = "upload\_here" \rightarrow \\
       & (decision = Grant)
\end{aligned}
\end{equation*}

and

\begin{equation*}
\begin{aligned}
pl_2': & (MEMBER = "data\_uploaders@example.com" \mid "jane@example.com" \mid \\
       & "bob@example.com" \mid "harry@example.com") \& \\
       & ROLE = "roles/storage.objectCreator" \& \\ 
       & (PERMISSION = prms_1 \mid \dots \mid prms_n) \& \\
       & RESOURCE = "upload\_here" \rightarrow \\
       & (decision = Grant)
\end{aligned}
\end{equation*}

Ultimately, the effective POLICIES at $project\_a$ and $upload\_here$ will be the union of the POLICIES directly applied to them and the POLICIES inherited from their ancestors. 

Table~\ref{ref:tbl2} shows the ROLES assigned to each MEMBER per RESOURCE. 

\begin{table}[t!]
\caption{Example 2 - Authorized roles per member and resource.}
\label{ref:tbl2}
\centering
\begin{tabular}{lcccc}
\hline
\multirow{3}{*}{Resource}          & \multicolumn{4}{c}{Authorized role}                                                                      \\ \cline{2-5} 
                                   & \multicolumn{1}{l}{\multirow{2}{*}{alice@example.com}} & \multicolumn{3}{c}{data\_uploaders@example.com} \\ \cline{3-5} 
                                   & \multicolumn{1}{l}{}                                   & jane           & harry           & bob          \\ \hline \hline
\multicolumn{1}{|l|}{example.com}  & \multicolumn{1}{c|}{No access}                         & \multicolumn{3}{c|}{No access}                  \\ \hline
\multicolumn{1}{|l|}{project\_a}   & \multicolumn{1}{c|}{Storage Object Admin}              & \multicolumn{3}{c|}{Storage Object Creator}     \\ \hline
\multicolumn{1}{|l|}{upload\_here} & \multicolumn{1}{c|}{Storage Object Admin}              & \multicolumn{3}{c|}{Storage Object Creator}     \\ \hline
\end{tabular}
\end{table}

The respective NuSMV code for the POLICIES and the transition system is available on GitHub \cite{github2020}, under Example 2. 
After expressing POLICIES and the $TS$ of the $RBAC_{GCP}$ in NuSMV, we specify the policy properties to be verified by the model checker. The following explains the evaluation of specifications.

Four of the example properties will be evaluated to be FALSE as follows.

\begin{verbatim}
SPEC AG ((MEMBER = "data_uploaders@example.com") & 
    (ROLE = ANY) & (PERMISSION = "storage.objects.delete") & 
    (RESOURCE = ANY) -> AF decision = Grant) 
\end{verbatim}

This property is FALSE since the group of $data\_uploaders@example.com$ does not have the permission $storage.objects.delete$ on any RESOURCE.

\begin{verbatim}
SPEC AG ((MEMBER = "alice@example.com") & 
    (ROLE = ANY) & (PERMISSION = ANY) & 
    (RESOURCE = "example.com") -> AF decision = Grant) 
\end{verbatim}

This property was also evaluated to be FALSE since we assigned $alice@example.com$ to the Storage Object Admin ROLE on $project\_a$, and from the RESOURCE hierarchy, it has no access on $example.com$ in a higher level.

\begin{verbatim}
SPEC AG ((MEMBER = ANY) & (ROLE = ANY) & 
    (PERMISSION = "storage.objects.delete" | 
    PERMISSION = "storage.objects.update" ) & 
    (RESOURCE = "example.com") -> AF decision = Grant) 
\end{verbatim}

The above property is evaluated to FALSE since, according to $RBAC_{GCP}$ RESOURCE hierarchy, none of the MEMBERS have the PERMISSION $storage.objects.delete$ or $storage.objects.update$ on $example.com$ because we assigned them to $project\_a$ that resides in a lower level.

\begin{verbatim}
SPEC AG ((MEMBER != "alice@example.com") & 
    (ROLE = ANY) & (PERMISSION = "storage.objects.create") & 
    (RESOURCE = ANY) -> AF decision = Deny) 
\end{verbatim}

This property is also FALSE since MEMBERS (different than  $alice@example.com$) have the PERMISSION $storage.objects.create$ on a RESOURCE at the RESOURCE hierarchy level. Group MEMBERS $data\_uploaders@example.com$ have PERMISSION for its assignment for the Storage Object Creator ROLE on $project\_a$ that resides at a higher level.

The verification of the first three specifications result is $RBAC.decision = Deny$ without a next state, which indicates that these properties can never be satisfied in the $RBAC_{GCP}$ model. The model checker NuSMV could not find any system state where the property would be evaluated to be TRUE so that it could eventually cause the access permission $Grant$ to happen. Similarly, the verification of the fourth specification results in $RBAC.decision = Grant$; hence, it is invalidated too.

\subsection{Example 3: Compute Engine}

The third example \cite{google2020} uses Compute Engine RESOURCES, which are virtual machines (VM) hosted on Google's infrastructure. For this example, the organization $example.com$, owns two projects, $project\_1$ and $project\_2$. And RESOURCES  $instance\_a$ and $instance\_b$ belong to each project respectively, as illustrated in Figure~\ref{fig:7}. Assuming that $bob@example.com$ is a MEMBER of the administrator's team that manages the network and security RESOURCES of the Organization, and $alice@example.com$ is a MEMBER of the development team. $bob@example.com$ is capable of making changes to all network RESOURCES and any project under it, and $alice@example.com$ should be allowed to launch instances and carry out other actions related to instances related to her project. Such security requirements are implemented by the POLICY on $example.com$ that assigns the Compute Network Admin ROLE ($roles/compute.networkAdmin$) to $bob@example.com$ and a second POLICY on $project\_2$ that assigns the Compute Instance Admin ROLE ($roles/compute.instanceAdmin$) to $alice@example.com$. The two POLICIES are:

\begin{figure}[!t]
  \centering
  \includegraphics[width=\linewidth]{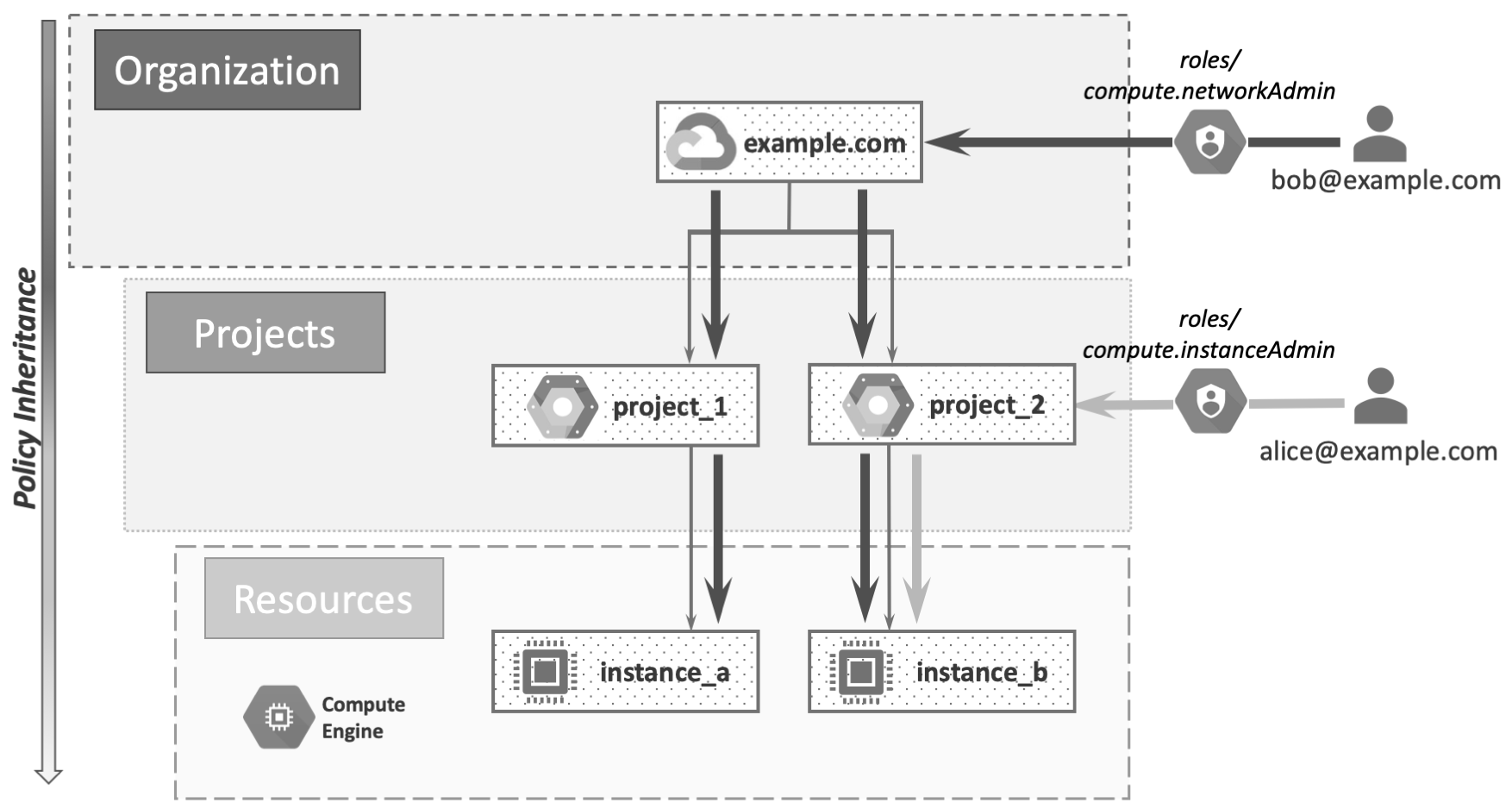}
  \caption{Example 3 - Compute Engine (based on~\cite{google2020}).}
  \label{fig:7}
\end{figure} 

POLICY for \textit{bob@example.com}:

\begin{equation*}
\begin{aligned}
pl_1: & MEMBER = "bob@example.com" \& ROLE = "compute.networkAdmin" \& \\
      & (PERMISSION = prms_1 \mid \dots \mid prms_n) \& RESOURCE = "example.com" \rightarrow \\
      & (decision = Grant)
\end{aligned}
\end{equation*}

POLICY for \textit{alice@example.com}:

\begin{equation*}
\begin{aligned}
pl_2: & MEMBER = "alice@example.com" \& \\
      & ROLE = "roles/compute.instanceAdmin" \& \\
      & (PERMISSION = prms_1 \mid \dots \mid prms_n) \& \\
      & RESOURCE = "project\_2" \rightarrow \\
      & (decision = Grant)
\end{aligned}
\end{equation*}

Since $project\_1$ and $project\_2$ inherit the POLICIES of $example.com$, once we define POLICY $pl_1$, we introduce the following POLICIES for the Compute Network Admin ROLE ($roles/compute.networkAdmin$) to be assigned to $bob@example.com$ on $project\_1$ and $project\_2$, as follows:

POLICY for \textit{bob@example.com} on $project\_1$:

\begin{equation*}
\begin{aligned}
pl_{1.1}: & MEMBER = "bob@example.com" \& \\
          & ROLE = "roles/compute.networkAdmin" \& \\
          & (PERMISSION = prms_1 | \dots | prms_n) \& \\
          & RESOURCE = "project\_1" \rightarrow \\
          & (decision = Grant)
\end{aligned}
\end{equation*}

POLICY for \textit{bob@example.com} on $project\_2$: 

\begin{equation*}
\begin{aligned}
pl_{1.2}: & MEMBER = "bob@example.com" \& \\
          & ROLE = "roles/compute.networkAdmin" \& \\
          & (PERMISSION = prms_1 \mid \dots \mid prms_n) \& \\
          & RESOURCE = "project\_2" \rightarrow \\
          & (decision = Grant)
\end{aligned}
\end{equation*}

RESOURCES $instance\_a$ and $instance\_b$ also inherit their parent resources' POLICY $project\_1$ and $project\_2$, respectively. The Compute Network Admin ROLE ($roles/compute.networkAdmin$) is assigned to $bob@example.com$ on $instance\_a$ and $instance\_b$, and the Compute Instance Admin ROLE ($roles/compute.instanceAdmin$) is assigned to $alice@example.com$ only on $instance\_b$. The introduced POLICIES are:

POLICY for \textit{bob@example.com} on $instance\_a$: 

\begin{equation*}
\begin{aligned}
pl_{1.1}': & MEMBER = "bob@example.com" \& \\
           & ROLE = "roles/compute.networkAdmin" \& \\
           & (PERMISSION = prms_1 | \dots | prms_n) \& \\
           & RESOURCE = "instance\_a" \rightarrow \\
           & (decision = Grant)
\end{aligned}
\end{equation*}

POLICY for \textit{bob@example.com} on $instance\_b$: 

\begin{equation*}
\begin{aligned}
pl_{1.2}': & MEMBER = "bob@example.com" \& \\
           & ROLE = "compute.networkAdmin" \& \\
           & (PERMISSION = prms_1 \mid \dots \mid prms_n) \& \\
           & RESOURCE = "instance\_b" \rightarrow \\
           & (decision = Grant)
\end{aligned}
\end{equation*}

POLICY for \textit{alice@example.com} on $instance\_b$: 

\begin{equation*}
\begin{aligned}
pl_2': & MEMBER = "alice@example.com" \& \\
       & ROLE = "roles/compute.instanceAdmin" \& \\
       & (PERMISSION = prms_1 \mid \dots \mid prms_n) \& \\
       & RESOURCE = "instance\_b" \rightarrow \\
       & (decision = Grant)
\end{aligned}
\end{equation*}

Ultimately, the effective POLICIES on every RESOURCE are the union of the POLICIES directly applied to the RESOURCE and the POLICIES inherited from its ancestors. 

Table~\ref{ref:tbl3} shows the ROLES assigned to each MEMBER per RESOURCE.

\begin{table*}[t!]

\caption{Example 3 - Authorized roles per member and resource}
\label{ref:tbl3}
\centering
\begin{tabular}{lcc}

\multicolumn{1}{c}{\multirow{2}{*}{Resource}} & \multicolumn{2}{c}{Authorized role}                                                      \\ \cline{2-3} 
\multicolumn{1}{c}{}                          & \multicolumn{1}{c|}{bob@example.com}       & alice@example.com                           \\ \hline
\multicolumn{1}{|l|}{example.com}             & \multicolumn{1}{c|}{Compute Network Admin} & \multicolumn{1}{c|}{No access}              \\ \hline
\multicolumn{1}{|l|}{$project\_1$}            & \multicolumn{1}{c|}{Compute Network Admin} & \multicolumn{1}{c|}{No access}              \\ \hline
\multicolumn{1}{|l|}{$project\_2$}            & \multicolumn{1}{c|}{Compute Network Admin} & \multicolumn{1}{c|}{Compute Instance Admin} \\ \hline
\multicolumn{1}{|l|}{$instance\_a$}           & \multicolumn{1}{c|}{Compute Network Admin} & \multicolumn{1}{c|}{No access}              \\ \hline
\multicolumn{1}{|l|}{$instance\_b$}            & \multicolumn{1}{l|}{Compute Network Admin} & \multicolumn{1}{l|}{Compute Instance Admin} \\ \hline

\end{tabular}
\end{table*}

The NuSMV code for the properties specification of this example is available on GitHub \cite{github2020}, under Example 3.

\begin{verbatim}
SPEC AG ((MEMBER = "alice@example.com") & 
    (ROLE = ANY) & 
    (PERMISSION = "compute.instances.create") & 
    (RESOURCE = "project_1") -> AF decision = Grant)
\end{verbatim}

This property will be evaluated to FALSE since $alice@example.com$ has the PERMISSION $compute.instances.create$ assigned to the Compute Instance Admin ROLE, but not on $project\_1$ that resides in a different branch of the RESOURCE hierarchy. ROLES do not affect peer RESOURCES.

\begin{verbatim}
SPEC AG ((MEMBER = ANY) & 
    (ROLE = "roles/compute.instanceAdmin") & 
    (PERMISSION = ANY) & (RESOURCE = "instance_a") -> AF decision = Grant)
\end{verbatim}

This property is FALSE since the Compute Instance Admin ROLE is assigned to $instance\_b$.

\begin{verbatim}
SPEC AG ((MEMBER = ANY) & (ROLE = ANY) & 
    (PERMISSION = "compute.instances.create") & 
    (RESOURCE = "project_1") -> AF decision = Grant)
\end{verbatim}

We have that $bob@example.com$ has access to $project\_1$, but his ROLE (Compute Network Admin) does not contain that specific PERMISSION. And $alice@example.com$ has this PERMISSION because of her assigned ROLE (Compute Instance Admin) on $project\_2$, but not on $project\_1$ in a different branch of the RESOURCE hierarchy. Hence, the property is FALSE since no one has the PERMISSION $compute.instances.create$ on $project\_1$.

In all three specifications, the result of the verification is $RBAC.decision = Deny$, without a next state since the NuSMV model checker could not find any system state where the properties is TRUE.

\subsection{Summary of Examples}
The first example used Cloud Pub/Sub RESOURCES and presented a case of RESOURCE hierarchy between a Project and a topic. We considered two different POLICIES for two MEMBERS, one on each RESOURCE. This example demonstrates how the $TS$ operates, and how properties are specified to check whether the hierarchy was implemented correctly. The second example used Cloud Storage RESOURCES to demonstrate the enforcement of two different POLICIES for two MEMBERS on the same RESOURCE. One of the MEMBERS is a Google group account that allowed us to investigate how the applied technique handles this type of a MEMBER. Google groups are a convenient way to apply organization access control policies and a best practice for role distribution \cite{google2020}. The third example used Compute Engine RESOURCES, which allowed us to evaluate the security policies in a more complex configuration where the resource structure contains more branches and nodes. Various properties in each example were checked to satisfy specific security requirements in compliance with Google's proposed best practices \cite{google2020}. Overall, the applied technique successfully verified the properties in all three examples; hence, offering the capability of a tool for administrators to specify policies/properties and verify their correctness.

\section{Conclusion}
\label{sec:6}

When defining policies in Cloud systems, it is imperative to understand the underlying access control model and supported policies to avoid configuration errors or even inconsistencies. Towards achieving this aim, we defined $RBAC_{GCP}$ to provide a better understanding of the RBAC model and policies supported by the Google Cloud IAM platform. The RBAC access control model of Cloud IAM has a few differences compared to the ANSI standard model. Specifically, the former supports permission inheritance through RESOURCE hierarchies but not ROLE hierarchies. We applied model checking to formally verify supported access control policies. And we demonstrated the technique's applicability through three examples described on the official Google Cloud IAM website. We anticipate this work to assist system administrators in ensuring the correctness of policy specification and checking violations against security requirement~\cite{gouglidis2014security} and, even further, performing a security assessment of policies for compliance purposes \cite{fcw2019}.

\section*{Acknowledgement} The authors would like to thank Dr Andrew Sogokon at Lancaster University for his feedback. This research is supported in part by the Security Lancaster VERIFi Mini-Project under grant number IRL1025.

%% If you have bibdatabase file and want bibtex to generate the
%% bibitems, please use
%%
 \bibliographystyle{elsarticle-num} 
%% \bibliography{main}

%% else use the following coding to input the bibitems directly in the
%% TeX file.

% \begin{thebibliography}{00}

% %% \bibitem{label}
% %% Text of bibliographic item

% \bibitem{}

% \end{thebibliography}
\end{document}